\documentclass[sn-mathphys,Numbered]{sn-jnl}

\usepackage{graphicx}%
\usepackage{multirow}%
\usepackage{amsmath,amssymb,amsfonts}%
\usepackage{amsthm}%
\usepackage{mathrsfs}%
\usepackage[title]{appendix}%
\usepackage{xcolor}%
\usepackage{textcomp}%
\usepackage{manyfoot}%
\usepackage{booktabs}%
\usepackage{algorithm}%
\usepackage{algorithmicx}%
\usepackage{algpseudocode}%
\usepackage{listings}%
\usepackage{siunitx}

\begin{document}

\title{Proton irradiation of plastic scintillator bars for POLAR-2}


\author*[1]{\fnm{Slawomir} \sur{Mianowski}}\email{slawomir.mianowski@ncbj.gov.pl}
\author*[2]{\fnm{Nicolas} \sur{De Angelis}}\email{nicolas.deangelis@unige.ch}
\author[1]{\fnm{Kamil} \sur{Brylew}}
\author[2]{\fnm{Johannes} \sur{Hulsman}}
\author[3]{\fnm{Tomasz} \sur{Kowalski}}
\author[3]{\fnm{Sebastian} \sur{Kusyk}}
\author[1]{\fnm{Zuzanna} \sur{Mianowska}}
\author[3]{\fnm{Jerzy} \sur{Mietelski}}
\author[1]{\fnm{Dominik} \sur{Rybka}}
\author[3]{\fnm{Jan} \sur{Swakon}}
\author[3]{\fnm{Damian} \sur{Wrobel}}

\affil*[1]{\orgname{National Centre for Nuclear Research}, \orgaddress{\street{A. Soltana 7 Street}, \city{Otwock}, \postcode{PL-05400}, \country{Poland}}}

\affil*[2]{\orgdiv{DPNC}, \orgname{University of Geneva}, \orgaddress{\street{24 Quai Ernest-Ansermet}, \city{Geneva}, \postcode{CH-1205}, \country{Switzerland}}}

\affil[3]{\orgname{Institute of Nuclear Physics Polish Academy of Sciences}, \orgaddress{\street{Radzikowskiego 152 Street}, \city{Krakow}, \postcode{PL-31342}, \country{Poland}}}

\abstract{
POLAR-2, a plastic scintillator based Compton polarimeter, is currently under development and planned for a launch to the China Space Station in 2025. It is intended to shed a new light on our understanding of Gamma-Ray Bursts by performing high precision polarization measurements of their prompt emission. The instrument will be orbiting at an average altitude of \SI{383}{km} with an inclination of \SI{42}{^\circ} and will be subject to background radiation from cosmic rays and solar events. 

In this work, we tested the performance of plastic scintillation bars, EJ-200 and EJ-248M from Eljen Technology, under space-like conditions, that were chosen as possible candidates for POLAR-2. Both scintillator types were irradiated with \SI{58}{MeV} protons at several doses from \SI{1.89}{Gy}(corresponding to about 13 years in space for POLAR-2) up to \SI{18.7}{Gy}, that goes far beyond the expected POLAR-2 life time. Their respective properties, expressed in terms of light yield, emission and absorption spectra, and activation analysis due to proton irradiation are discussed.  
Scintillators activation analyses showed a dominant contribution of $\beta^{+}$ decay with a typical for this process gamma-ray energy line of \SI{511}{keV}.
}

\keywords{POLAR-2, plastic scintillator, radiation, protons, cosmic rays}



\maketitle

\section{Introduction}\label{sec:intro}

POLAR-2 is a space-borne polarimeter, that will be launched to the China
Space Station (CSS) in 2025 for a mission of at least 2 years. The CSS (and hence POLAR-2) is orbiting at a typical altitude of \SI{383}{km} with an inclination of \SI{42}{^\circ}. As a result, it is exposed to radiation from cosmic rays of galactic, solar and trapped origin \cite{POLAR-2_SiPMirradiation}.

\begin{figure}[!h]
  \centering
  \includegraphics[height=0.5\textwidth]{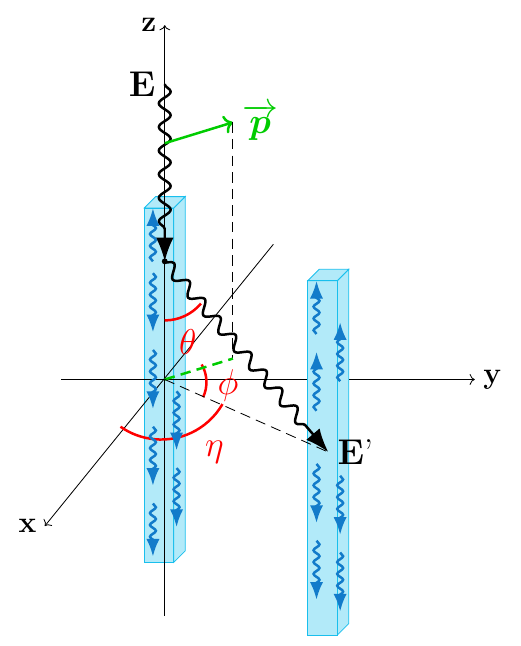}\hspace*{0.5cm}\includegraphics[height=0.5\textwidth]{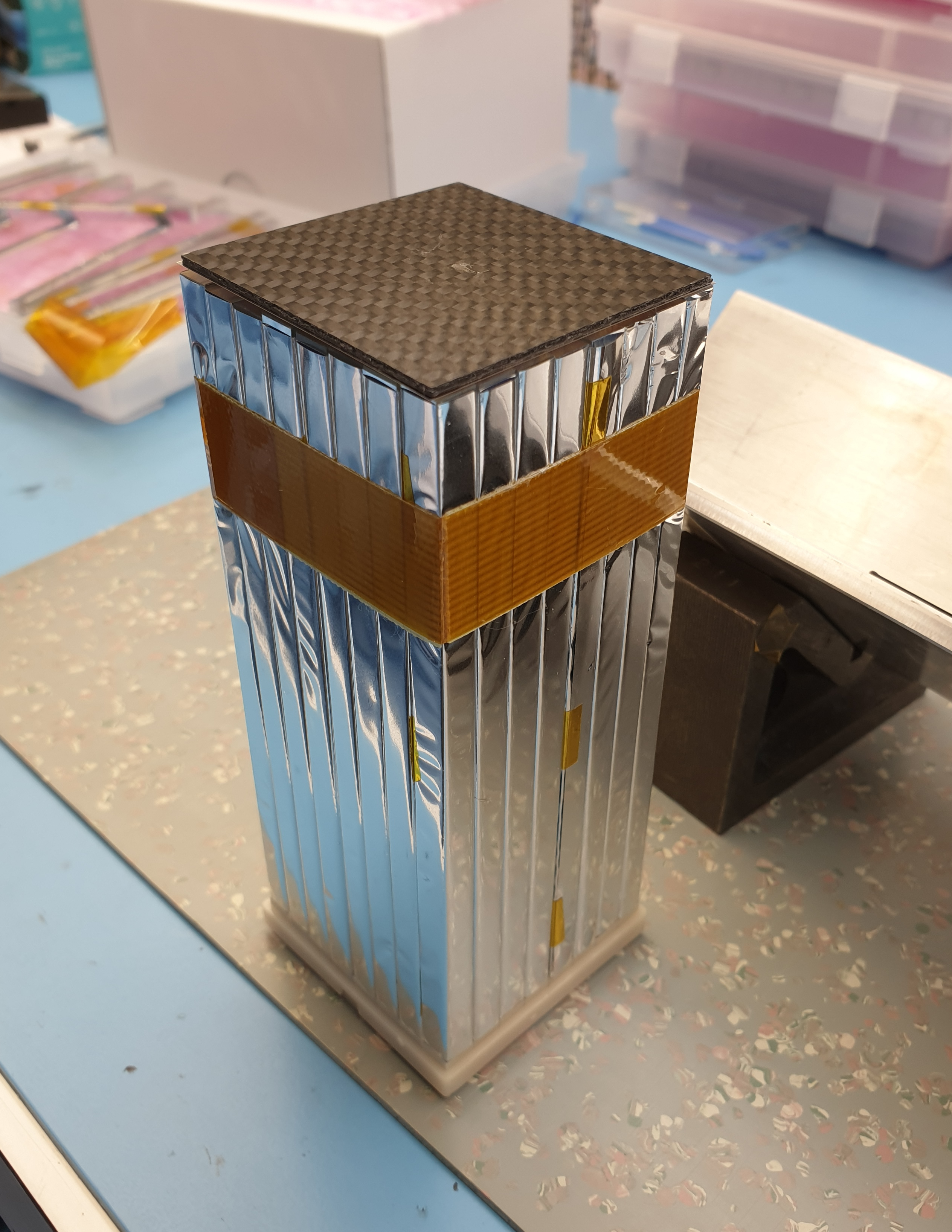}
  \caption{Left - Compton scattering of an incoming $\gamma$-ray between two scintillator bars \cite{ICRC21_POLAR-2_NDA}. Right - Assembled POLAR-2 target made of sixty four EJ-248M bars wrapped in 3M\texttrademark ~Vikuiti and Toray Claryl\textregistered reflective foils.}
  \label{fig:scatt}
\end{figure}

The POLAR-2 detection principle is based on its predecessor mission POLAR \cite{ProduitPOLAR}. Some improvements have been made in order to lower the low energy threshold down to a few keV\footnote{This is the detection for a single bar event. The energy threshold for polarization measurements is reduced from \SI{50}{keV} down to \SI{30}{keV}, higher than the single bar threshold since the photon need to deposit energy in at least two bars for measuring polarization.}, like reducing the dead-space between channels with wider bars and optimized mechanics as well as upgrading the photosensors used to read out the scintillators from PhotoMultiplier Tubes (PMTs) to Silicon PhotoMultipliers (SiPMs) in order to improve the light yield of the overall system. The scintillator bars length has also been optimized to reduce the background contribution while still having decent statistics for typical Gamma-Ray Bursts (GRBs). The full polarimeter is composed of 100 modules, each made of 64 plastic scintillators with dimensions\footnote{Note that the POLAR scintillator bars had dimensions of \SI{5.8}{mm}$\times$\SI{5.8}{mm}$\times$\SI{176}{mm}.} \SI{5.9}{mm}$\times$\SI{5.9}{mm}$\times$\SI{125}{mm} (see Figure \ref{fig:scatt}), resulting in 6400 plastic bars in total.

In the case of the POLAR-2 polarimeter, segmented into elongated scintillator bars, we want the $\gamma$ photons to Compton scatter in a first bar, and ideally be completely absorbed in a second bar. As depicted in Figure \ref{fig:scatt}, the scattering angle is linked to the polarization vector, as the photon preferentially scatters orthogonal to that vector. Detecting many scattering events will therefore lead to a scattering angle distribution which will provide information on the polarization parameters (fraction and angle). In order to optimize the instrument for Compton scattering down to a few keV, the polarimeter requires a low-Z material, which explains the choice of plastic scintillators.

POLAR-2's predecessor successfully used EJ-248M scintillator bars. In this paper we are comparing the optical properties of this scintillator with a new EJ-200, which as manufacturer specification says \cite{Eljen}, is characterized by a higher light yield and a longer light attenuation length. Space-like conditions were reproduced in the laboratory by dedicated 58~MeV proton irradiation sessions.

\section{Samples and irradiation setup}
\begin{figure}[!h]
  \centering
  \includegraphics[scale=0.13]{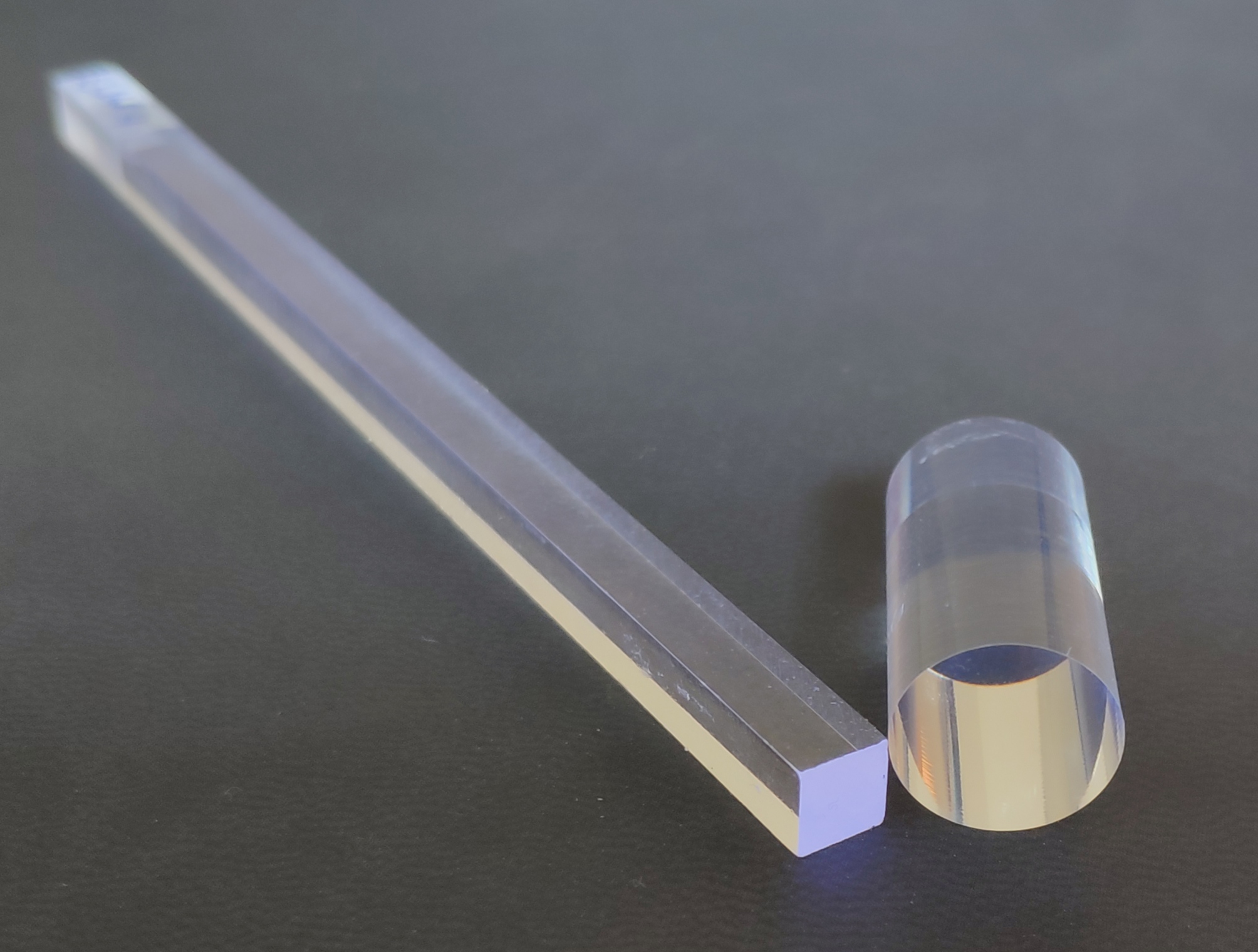}
  \caption{Two EJ-200 plastic scintillators. Left - scintillator bar chosen for POLAR-2. Right - cylindrical plastic as our reference point (see details in the text).}
  \label{fig:plastics}
\end{figure}

As it was presented in \cite{ICRC21_POLAR-2_NDA, ICRC21_POLAR-2_MK} and shown in Figure \ref{fig:scatt}, each of the 100 POLAR-2 detector modules consists of a target of 64 plastic scintillators. Figure \ref{fig:plastics} shows an example of two EJ-200 scintillators, which are visually indistinguishable from EJ-248M. The left one, a plastic bar with size \SI{5.9}{mm}$\times$\SI{5.9}{mm}$\times$\SI{125}{mm}, was chosen as a candidate for POLAR-2, while the one on the right, with a cylindrical shape and dimensions of $\Phi$\SI{12.7}{mm}$\times$\SI{25.4}{mm}, is used as our reference point since the elongated shape of the bar may affect the light yield.

EJ-200 plastic scintillator combines two important properties of long optical attenuation length and high scintillation efficiency, whereas the softening point is around 70$^{\circ}$C. EJ-248M has very similar properties except for a shorter attenuation length and a lower scintillation efficiency, while being sustaining higher temperatures. The higher softening temperature has been achieved by using a specially modified variant of the conventional PVT base plastic. Table \ref{tab:plastics} summarises basic properties of both types of scintillator based on information provided by the manufacturer \cite{Eljen}.

\begin{table}[ht]
\centering
\begin{tabular}{| c || c | c |}
\hline
 
Properties & EJ-200 & EJ-248M \\ 
\hline
Density (g/cm$^3$) & 1.023  & 1.023 \\
\hline
Max. emission (nm)  & 425 & 425 \\ 
\hline
Scint. efficiency (photons/keV) & 10 & 9.2 \\
\hline
Rise/decay time (ns) & 0.9/2.1 & 0.9/2.1  \\
\hline
Light Attenuation Length (cm) & 380 & 250 \\
\hline
Softening point ($^\circ$C) & 75   & 99 \\
\hline

\end{tabular}
\caption{Basic properties of EJ-200 and EJ-248M plastic scintillators, considered as candidates for the POLAR-2 polarimeter \cite{Eljen}.}
\label{tab:plastics}
\end{table}

\subsection{Scintillators proton irradiation}

All planned irradiation sessions were performed at the proton radiotherapy facility \cite{MICHALEC2010738} at the Institute of Nuclear Physics Polish Academy of Sciences (IFJ PAN) in Krakow. The irradiation campaign was planned in order to reproduce the integrated dose the POLAR-2 instrument will face during its lifetime in orbit (which includes cosmic radiation background and solar events as well as passages through the South Atlantic Anomaly). The total dose, activation and optical performances of the plastic bars are of interest. This analysis completes our previous studies, that were focused on silicon photomultipliers properties in radiation environments \cite{POLAR-2_SiPMirradiation}. Those photodetectors were chosen to read the light from the plastic scintillator bars, that are the subject of this paper.   

\begin{figure}[!h]
  \centering
  \includegraphics[width=0.8\textwidth]{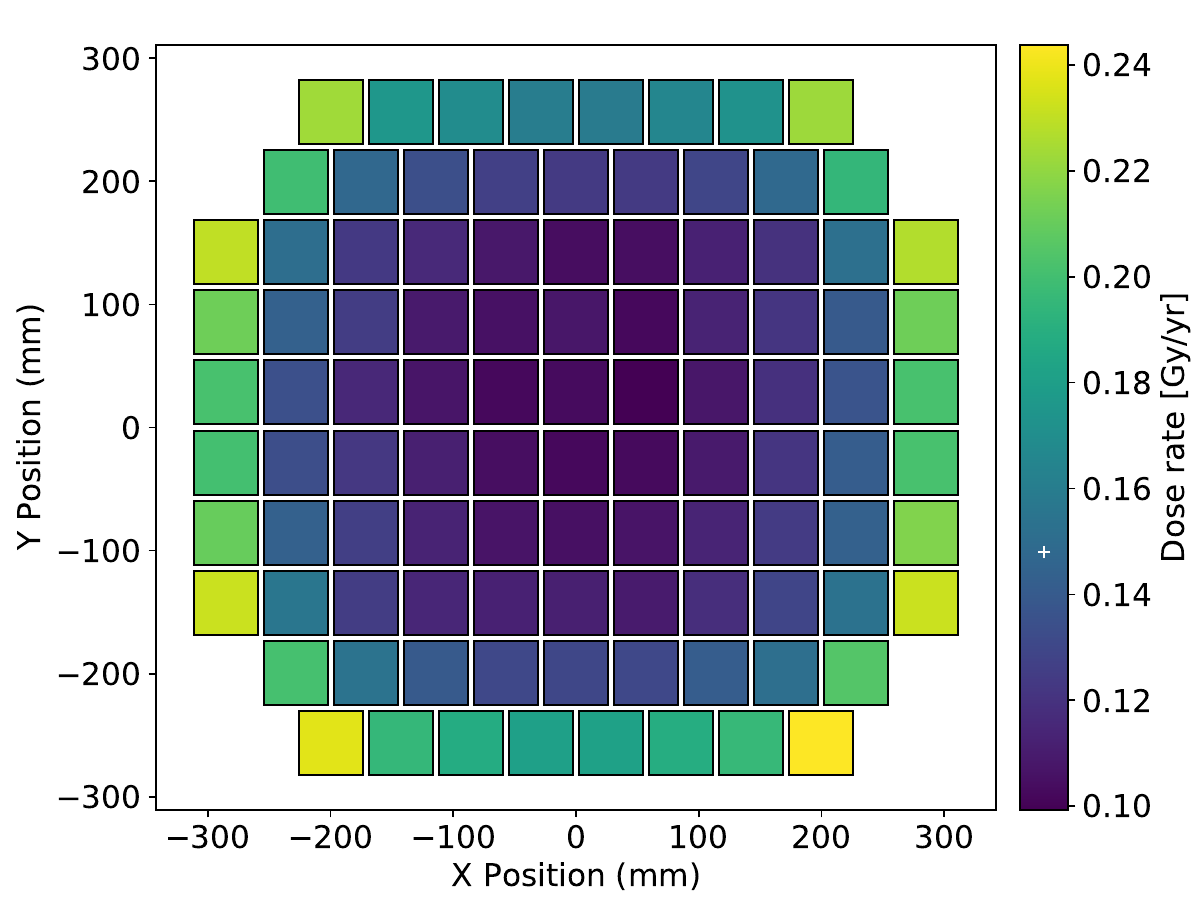}
  \caption{Scintillators dose rate distribution among the 100 POLAR-2 polarimeter modules. The white marker on the colorbar shows the dose rate averaged over the full instrument.}
  \label{fig:dose_map_scint}
\end{figure}

A schematic of this facility, a photo of the exit of the beam line and proton beam parameters were presented and more detailed described in \cite{POLAR-2_SiPMirradiation, POLAR-2Annealing}. The dose rate at which the scintillators are exposed in space, averaged over the entire polarimeter, were computed using LEOBackground \cite{leobackground_2019} and SPENVIS \cite{1989AIPC..186..483D} for modelling the radiation environment and Geant4 \cite{AGOSTINELLI2003250} for simulating the radiation deposition in the bars. A detailed description of the simulation setup is given in \cite{POLAR-2_SiPMirradiation}. Figure \ref{fig:dose_map_scint} shows the dose distribution among the POLAR-2 modules, the outside channels being more exposed to cosmic radiations. Two scenarios were considered to estimate the annual dose at which the plastic scintillators are exposed: the so-called 'Bare scintillator', where the scintillator is placed directly in space without any shielding, and a more realistic scenario referred as 'Full Instrument + CSS', where scintillator bars are shielded by the China Space Station (CSS) and other polarimeter elements (the entire POLAR-2 design is therefore implemented in this latter scenario). A dose rate of $2.38\times10^{-1}$~Gy/yr is expected for the 'Bare scintillator' scenario while the 'Full Instrument + CSS' scenario leads to a yearly dose of $1.48\times10^{-1}$~Gy. Table \ref{tab:rad_table} shows the list of irradiated plastic scintillator bar samples, their corresponding dose and equivalent time in space for a both scenarios.

\begin{table}[ht]
\centering
\begin{tabular}{ | c | c | c| c | c |}
\hline
Sample & Fluence & Total Dose & \multicolumn{2}{|c|}{Equivalent Years in Space for POLAR-2}\\ 
Type &($\frac{protons}{cm^2}$) & (Gy) & 'Full Instr. + CSS' & 'Bare scintillator' \\ 
\hline
EJ-200-1 and EJ-248M-1  & $1.00 \times 10^9$ & 1.89 & 12.8 & 7.94 \\ 
\hline
EJ-200-2 and EJ-248M-2  & $3.00 \times 10^9$ & 5.66 & 38.2 & 23.8 \\
\hline
EJ-200-3 and EJ-248M-3  & $5.92 \times 10^9$ & 11.2 & 75.7 & 47.1 \\ 
\hline
EJ-200-4 and EJ-248M-4  & $9.93 \times 10^9$ & 18.7 & 126 & 78.6 \\
\hline
\end{tabular}
\caption{An overview of irradiated plastic bars, their corresponding dose and equivalent time in space for a 'Full Instrument + CSS' and 'Bare plastic bar' scenarios.}
\label{tab:rad_table}
\end{table}

\newpage
\section{Radioluminescence emission and absorption spectra characterization}

As a first comparison point, radioluminescence emission (RL) spectra were measured for EJ-200 and EJ-248M scintillators before and after proton irradiation.

\begin{figure}[!h]
  \centering
  \includegraphics[scale=0.25]{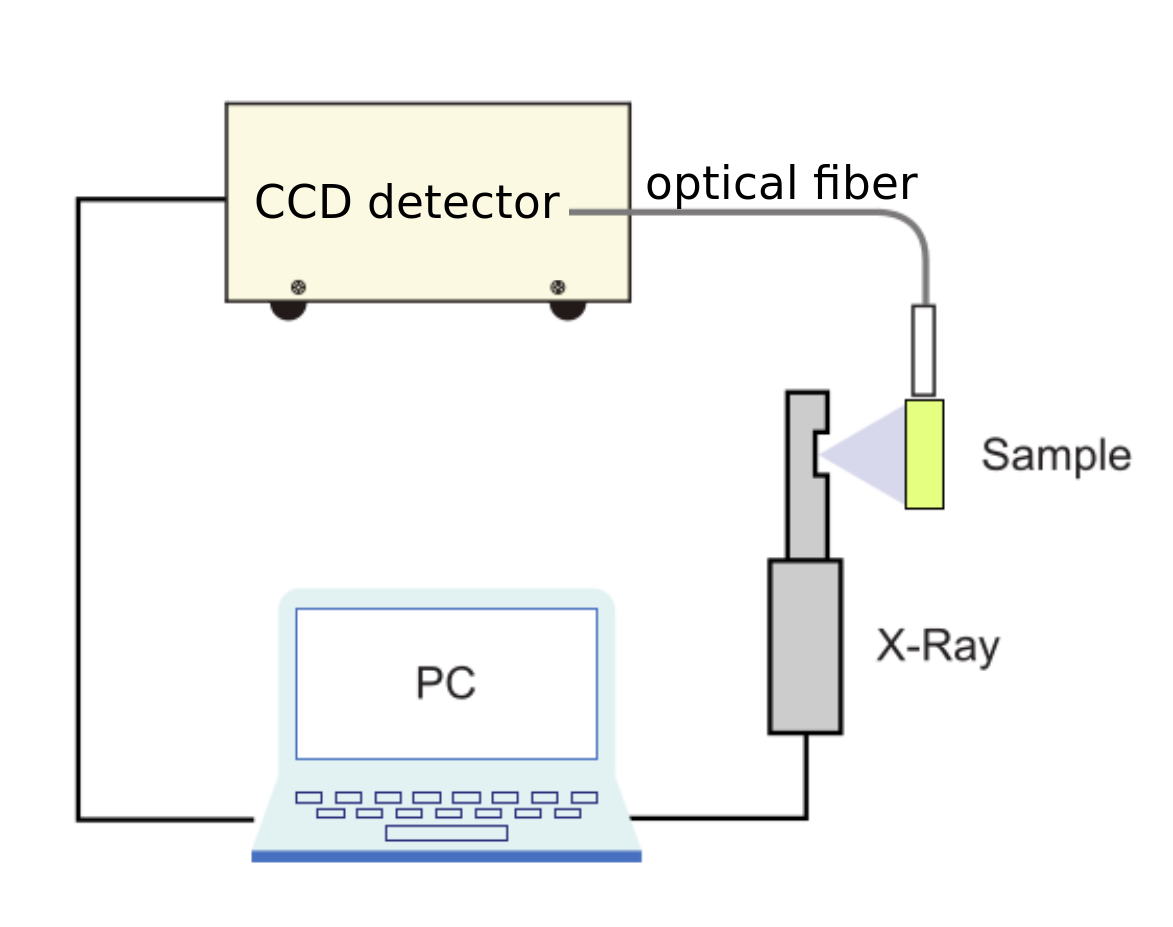}
  \caption{Experimental set-up for radioluminescence emission of plastic bars.}
  \label{fig:X-ray}
\end{figure}

\begin{figure}[!h]
  \centering
  \includegraphics[scale=1.1]{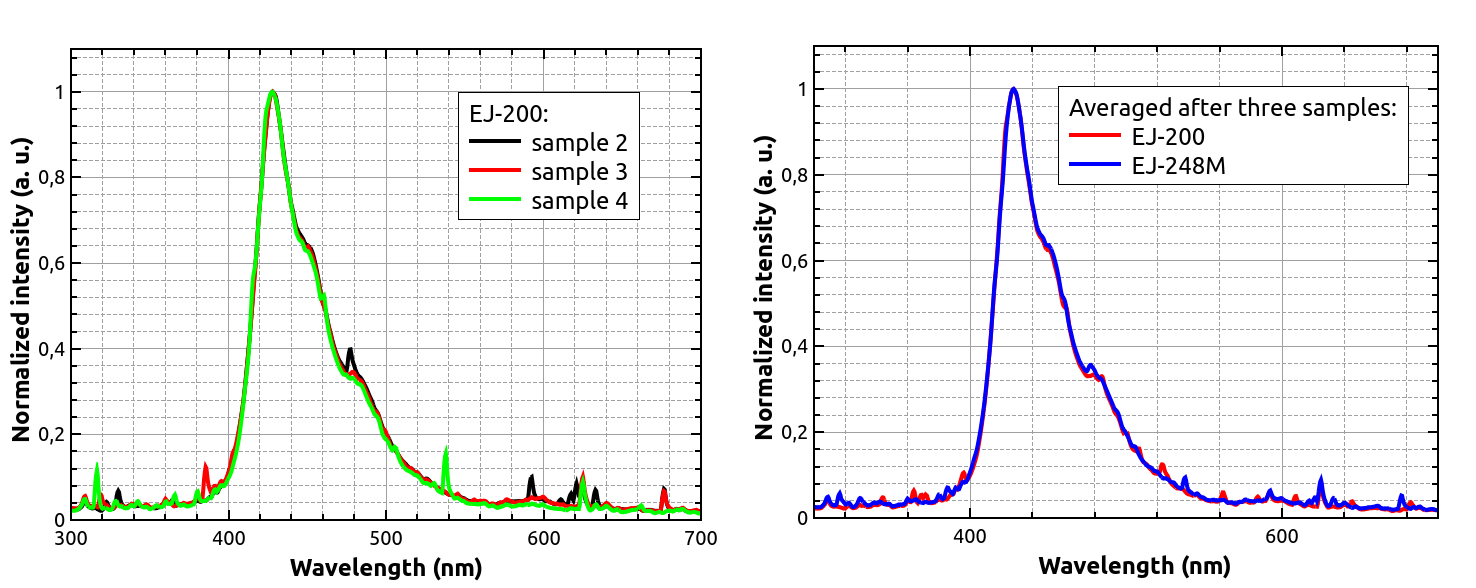}
  \caption{Comparison of radioluminescence emission spectrum of plastic bars. Left - three RL spectra for EJ-200 measured before irradiation. Right - averaged after three samples for each type of scintillator RL spectra for EJ-200 and EJ-248M before irradiation.}
  \label{fig:RL_spec_before}
\end{figure}

The RL excitation was performed using a MiniX Amp-Tek X-ray tube operating at 20~$\mu$A and 20~kV. The luminescence spectrum was registered using a Hamamatsu PMA-12 Photonic ccd analyzer measured in the range of 200~nm to 900~nm. The detector has a resolution lower than 2~nm. The emission spectrum was read using optical fiber from the front scintillator surface (5.9$\times$5.9~mm$^2$), while the X-ray tube and emitted X-ray photons were located perpendicular to that surface. The experimental set-up is presented in figure \ref{fig:X-ray}. The integration time was 60~ms and the spectra have been averaged after 10 repetitions. The RL spectra obtained for three different samples of EJ-200\footnote{One sample of each scintillator type was delivered later, since it was used for activation measurements right after irradiation (see Section \ref{sec:activation}).} and averaged spectra for EJ-200 and EJ-248M before proton irradiation are presented in figure  \ref{fig:RL_spec_before}. No significant variations of the emission spectrum are observed between samples of the same material, and neither between EJ-200 and EJ-248M, which agrees the manufacturer specification. Measurement uncertainties are up to 5$\%$ in the 350-550~nm range.

\begin{figure}[!h]
  \centering
  \includegraphics[scale=1.1]{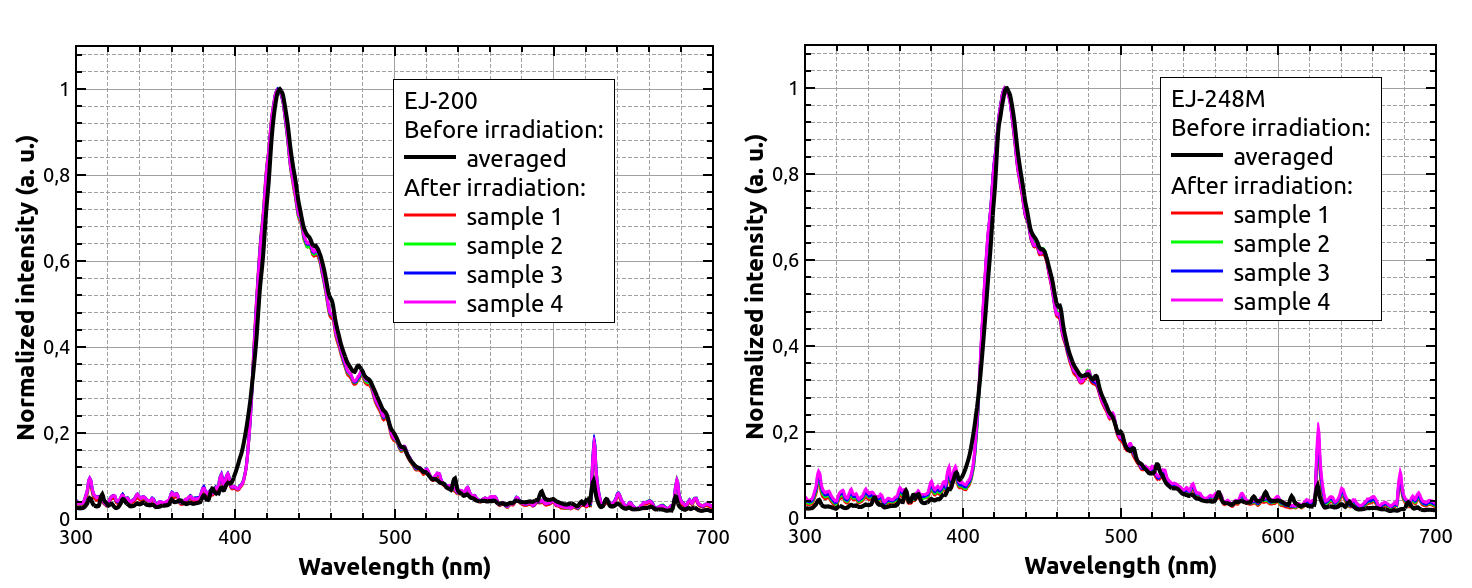}
  \caption{Comparison of radioluminescence emission spectrum of plastic bars. Left - RL spectra for EJ-200 measured after irradiation for samples 1-4. Right - RL spectra for EJ-248M measured after irradiation for samples 1-4. Black lines show averaged signals before irradiation (see Figure \ref{fig:RL_spec_before})}
  \label{fig:RL_spec_after}
\end{figure}

Data obtained for irradiated samples with doses presented in table \ref{tab:rad_table} are shown in figure \ref{fig:RL_spec_after}. No significant changes in the RL spectra were observed after irradiation, even for the highest dose. Statistical uncertainties on the measurement intensity are less than 5\%. A spectral deepening is observed after irradiation in the 400-410~nm interval.

\begin{figure}[!h]
  \centering
  \includegraphics[scale=0.35]{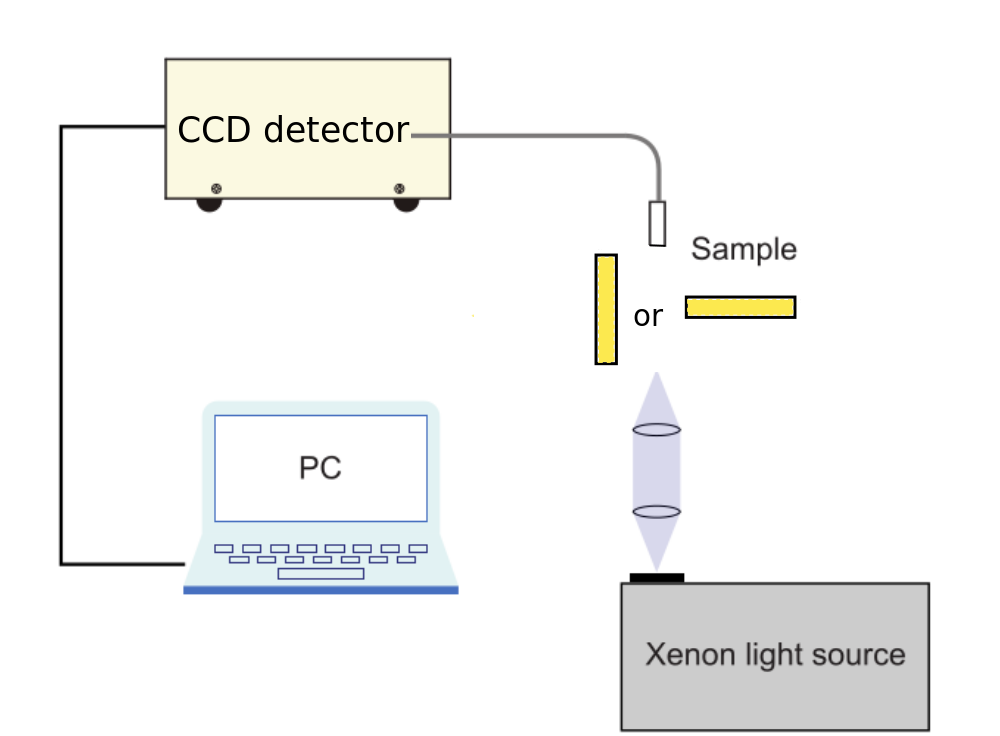}
  \caption{Experimental set-up for absorption measurements, where the light was transmitted along scintillator length.}
  \label{fig:abs_setup}
\end{figure}

As a second comparison point, the absorption spectra were measured after proton irradiation for both materials. A Spectral Products ASB-XE-175EX xenon lamp was used as a source of white light (Figure \ref{fig:abs_setup}), while the same Hamamatsu PMA-12 ccd detector previously mentioned was employed. The spectrum of the xenon lamp was measured unobstructed by the investigated sample as a reference measurement. Samples were then placed in the light beam and the spectrum was recorded once again.

\begin{figure}[!h]
  \centering
  \includegraphics[scale=1.1]{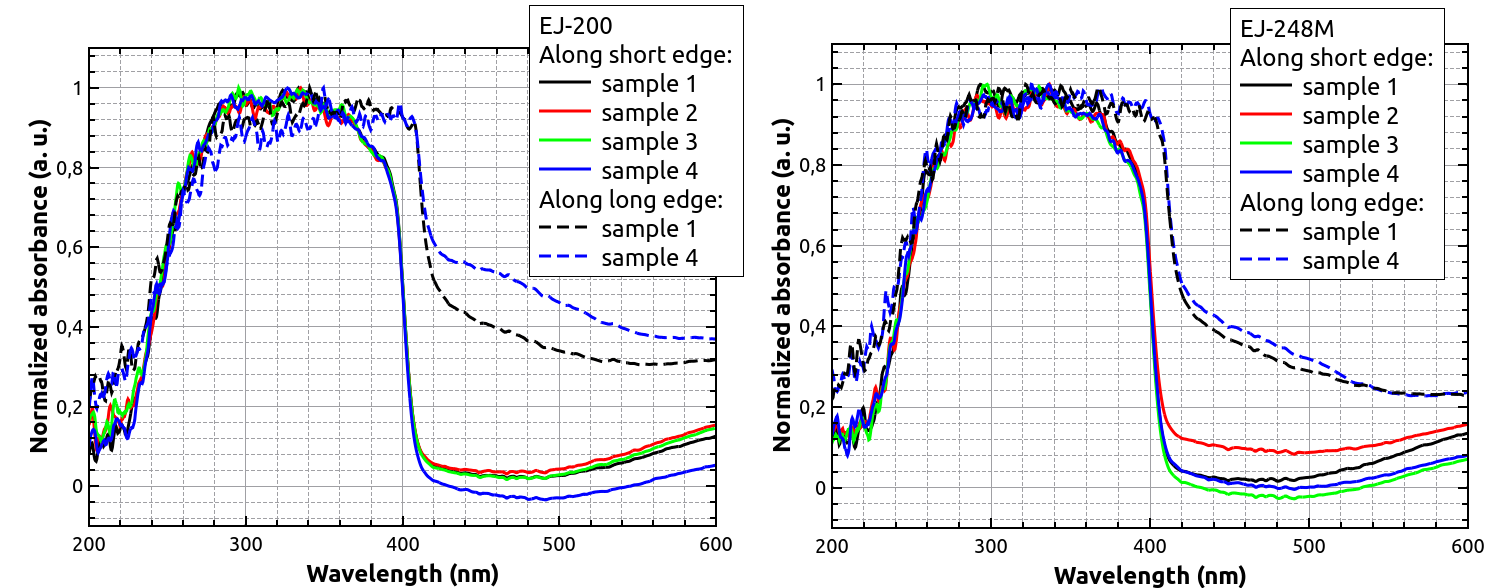}
  \caption{Absorbance measured for irradiated sampled of EJ-200 (left) and EJ-248M (right) scintillators. Solid lines correspond to the cases, where measured was done along short scintillator edge (5.9~mm) and dashed lines - along the longest scintillator edge.}
  \label{fig:absorbance}
\end{figure}

The absorption was measured both along shorter scintillator edge for all samples (with the samples placed perpendicularly to the light source), and along the scintillator length for samples 1 and 4 (the lowest and the highest dose). In each case, the wavelength-dependent absorbance (A) was calculated using the formula:
\begin{equation}
\textnormal{A}(\lambda) = -\textnormal{ln} \frac{\textnormal{I}(\lambda)}{\textnormal{I}_0(\lambda)},
\end{equation}
where I$_0$ is the initial beam intensity at a given wavelength $\lambda$ and I is the intensity of the beam after passing through the sample. The integration time was 100~ms and the measurements were repeated 60~times and then averaged. Because we could not ensure exactly the same geometry for each sample during the measurements, all spectra were normalized to unity, to enhance the potential the spectral shape changes. Figure \ref{fig:absorbance} shows the obtained results. As in the case of the radioluminescence measurements, we did not observe any significant changes between irradiated samples, and the highest absorption range is between about 250~nm and 400~nm (or up to 420~nm when measuring along the length of the bar).

In conclusion, both scintillator types show similar spectral properties. Emission and absorption spectra did not show significant changes due to irradiation with different doses. This confirms good radiation hardness properties of EJ-200 and EJ-248M for the expected POLAR-2 dose ranges.

\section{Position dependent scintillators' light yield measurement}

In this section we focused on optical light yield measurements as a function of the distance from the photodetector surface for both types of scintillators. For this purpose each scintillator bar was wrapped by high reflectivity 3M Vikuiti foil, that is planned to be used in the final POLAR-2 detector design. To optimize the light collection, the plastic bars were coupled to the photodetector using optical grease.

\begin{figure}[!h]
  \centering
  \includegraphics[height=0.42\textwidth]{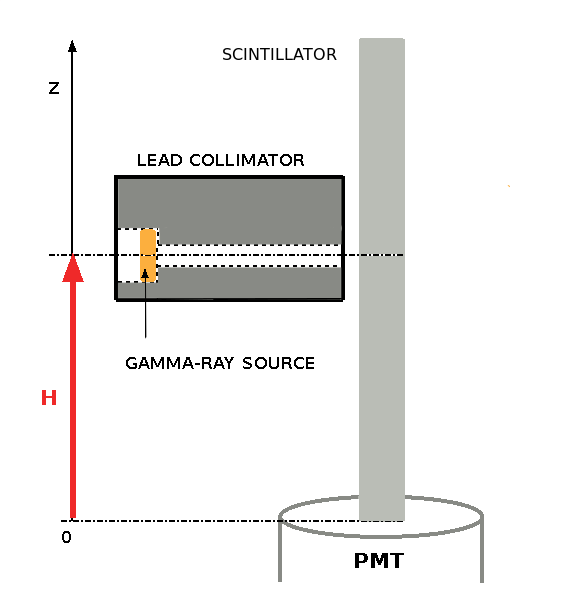}\includegraphics[height=0.42\textwidth]{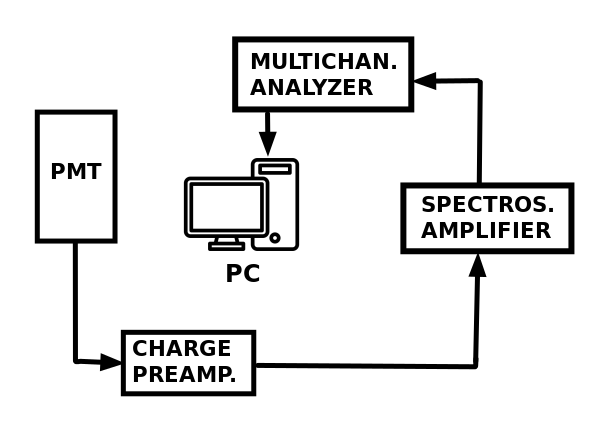}
  \caption{Experimental set-up scheme. Left - PMT, scintillator and collimator configuration. Right - analog electronic readout scheme.}
  \label{fig:PMT}
\end{figure}

\begin{figure}[!h]
  \centering
  \includegraphics[scale=1.4]{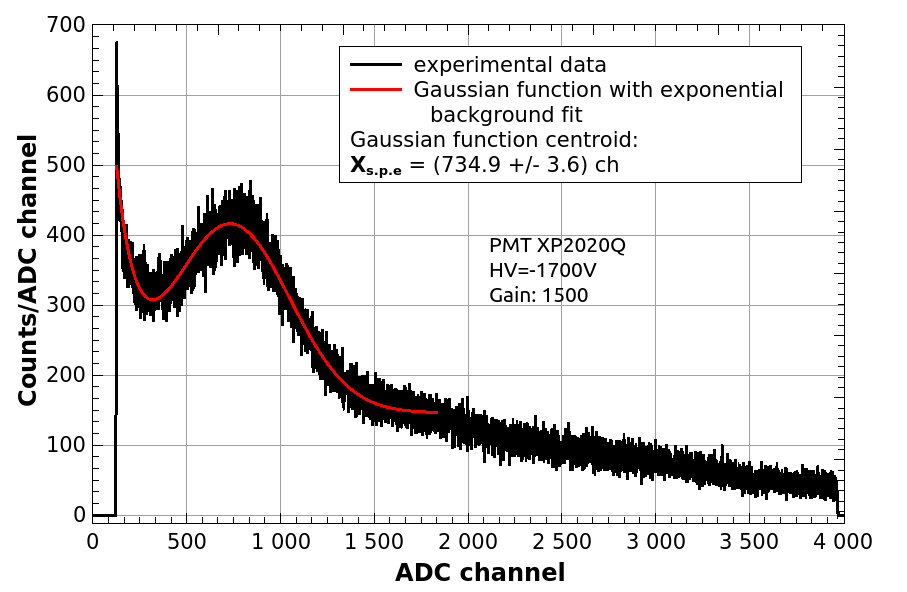}
  \caption{Example of dark count spectrum measured for PMT XP2020Q with single photo-electron peak. Red curve shows the Gaussian fit used to determine the peak position.}
  \label{fig:SPE}
\end{figure}

During those studies an electronic set-up based on analog readout was used. A calibrated classical photomultiplier (PMT) Photonis XP2020Q was chosen for light readout. This detector is characterized by very fast single photo-electron (p.e.) response (FWHM about 2.4~ns) and average quantum efficiency of about 25$\%$. The light-induced electrical pulse is sent from the voltage divider to a fast charge preamplifier followed by a spectroscopic amplifier. The shaped signal is finally recorded by a TUKAN 8K multichannel analyzer. This set-up allowed us to improve the signal-to-noise ratio and optimize the gain for measurements with a $^{137}$Cs $\gamma$-ray source. The same set-up was chosen to determine the dark count PMT spectrum and the position of the single p.e. peak, needed to determine the light yield. An example of dark count spectrum is presented in Figure \ref{fig:SPE}, where the single p.e. peak position is computed performing a two components function fit. The first component of the fit is a Gaussian function that characterizes the single p.e. peak, and the second component is an exponential function, describing the falling edge of the noise contribution, as can be seen at low ADC counts in the spectrum.

To determine the gamma-rays interaction point in a plastic scintillator, the $^{137}$Cs source was placed in a lead collimator (see Figure \ref{fig:PMT}). The collimator radius is \SI{3}{mm}, value which is used as the uncertainty on the height H above the PMT glass window.

\begin{figure}[!h]
  \centering
  \includegraphics[scale=1.1]{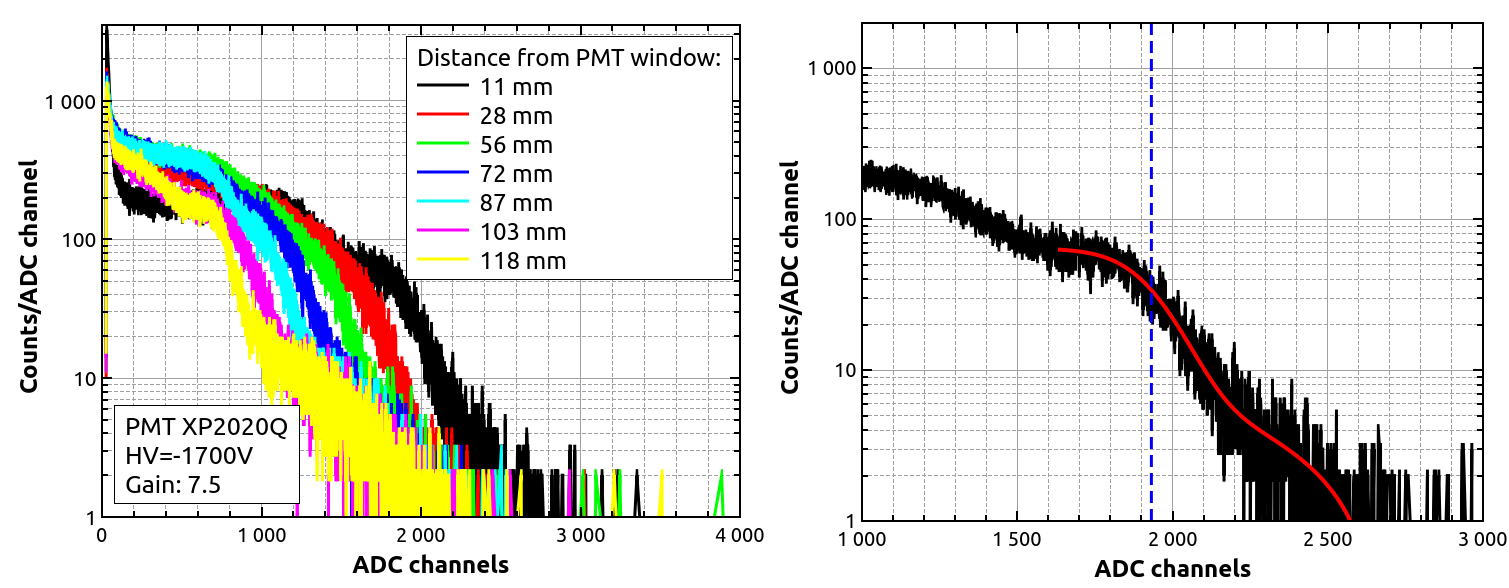}
  \caption{Energy spectrum of $^{137}$Cs measured for EJ-200 plastic bar: left - as a function of distance from PMT window, right - the position of Compton edge (blue dashed line) determined by the fitting equation \ref{eq:compton} (red line).}
  \label{fig:E_EJ200_noIrr}
\end{figure}

The light yield as a function of distance along the z-axis in the units of photo-electrons was calculated using following equation:
\begin{equation}
\textnormal{N}_{p.e.}(z) = \frac{\textnormal{X}_{\gamma}(z)}{\textnormal{X}_{s.p.e.}} \cdot \frac{\textnormal{G}_{s.p.e}}{\textnormal{G}_{\gamma}} ,
\end{equation}
where X$_{\gamma}(z)$ is the Compton edge position in the $^{137}$Cs $\gamma$-ray energy spectrum, X$_{s.p.e.}$ is a single photo-electron peak position, both in channels units (see figure \ref{fig:SPE}) and G$_{s.p.e}$, G$_{\gamma}$ are the respective gain values for the $\gamma$-ray energy spectrum and dark count spectrum. Figure \ref{fig:E_EJ200_noIrr}-left presents the $\gamma$-ray energy spectra measured for 662~keV gamma line from $^{137}$Cs and for EJ-200 plastic bar before proton irradiation for seven distances ranging from 11~mm to 118~mm from the PMT window. Due to the low scintillator density, only Compton edge region was clearly visible (the photo-peak at 662~keV was not detected). The Compton edge position, whose energy is around 476~keV, was determined each time by fitting with a sum of linear background and complementary error function (Erfc):
\begin{equation}
\textnormal{A}(ch, z) = \textnormal{A}_0+\textnormal{A}_1 \cdot ch+\textnormal{A}_2 \cdot \textnormal{Erfc}((ch-\textnormal{X}_{\gamma}(z))/\textnormal{A}_3),
\label{eq:compton}
\end{equation}
where A$_{0-3}$ are free fitting parameters. This parametrization determines the Compton edge position at half of the Compton valley height\footnote{Other parametrizations are also known, see for example \cite{Swiderski}, but they do not change the main conclusions.}. Figure \ref{fig:E_EJ200_noIrr}-right shows the Compton edge position based on this assumption.  

\begin{figure}[!h]
  \centering
        \includegraphics[scale=1.3]{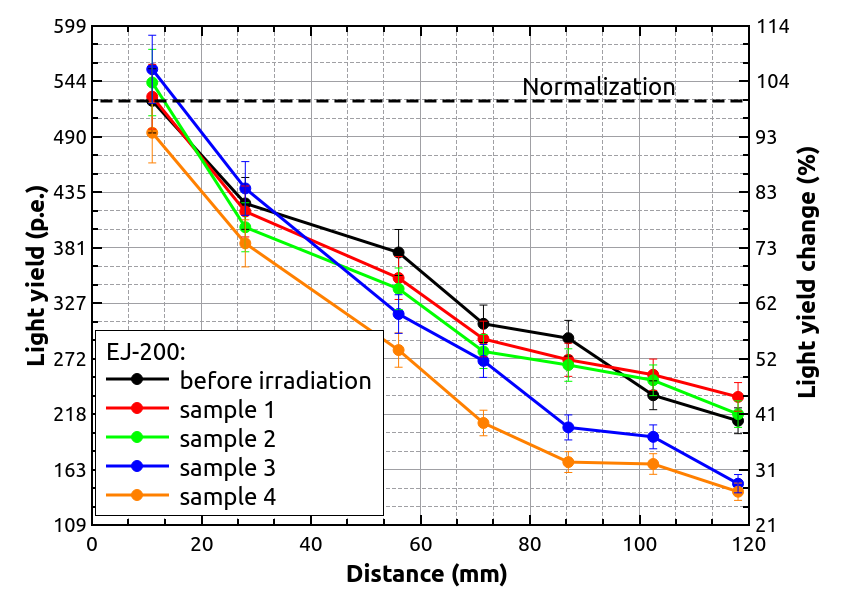}
    \centering
        \includegraphics[scale=1.3]{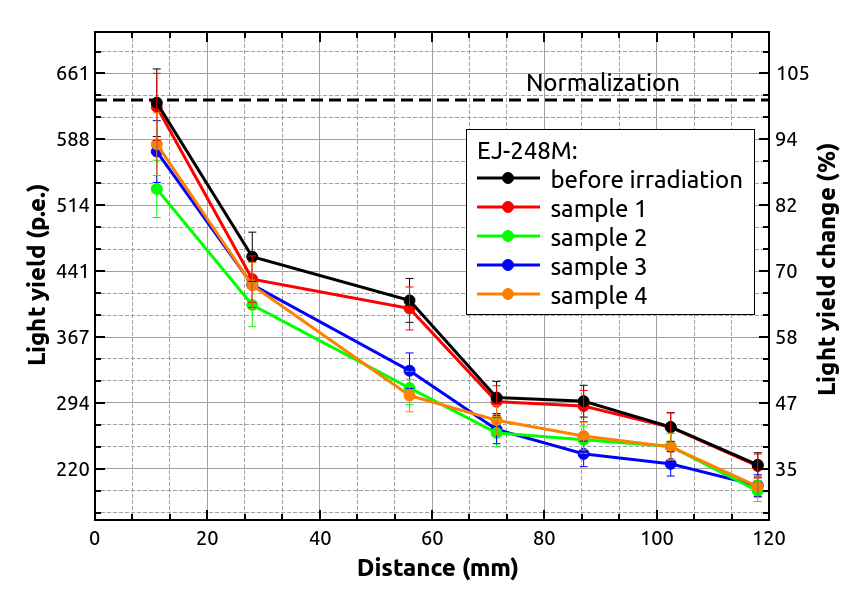}
  \caption{Light yield of EJ-200 (top) and EJ-248M (bottom) expressed in photo-electron units as a function of distance from PMT window for non-irradiated and irradiated samples. The light yield change normalization point was chosen as a value for non-irradiated scintillator at distance of 11~mm.}
  \label{fig:LY}
\end{figure}

Finally, figure \ref{fig:LY} shows the light yield as a function of the height over the PMT window for EJ-200 and EJ-248M scintillator bars before and after proton irradiation with doses up to 18.7~Gy. It can be seen that both types of scintillator have similar light yield characteristics. In general, the number of emitted photo-electrons is in the range of 500-600 p.e. for 476~keV at distance of 11~mm, and these values are reduced to about 30$\%$-40$\%$ at 118~mm. As it was showed for emission and absorption spectrum, we did not observe significant degradation with scintillator dose increase. Observed differences may be explained, in our opinion, by calculated uncertainties (6$\%$-7$\%$) and by more technical aspects like scintillators wrapping and coupling to the PMT, and different techniques of surface polishing \footnote{Manufacturer private communication. The relation of the light yield with the scintillator polishing quality will be investigated in a future publication about POLAR-2 module optical simulations and characterization \cite{POLAR2Optsim}.}. To confirm this observation, the same scintillators type but with cylindrical shape ($\phi$ 0.5~inch$\times$1~inch) were measured in the same way. This time the $^{137}$Cs source was placed directly on the sample top. Obtained values: (970$\pm$48)~p.e. for non-irradiated EJ-200 and (913$\pm$56)~p.e. for non-irradiated EJ-248M seem to confirm our expectations and obtained uncertainty ranges. At the same time, these results show, that geometrical shape expressed by the scintillator volume to surface area ratio (cylindrical/bar = 3.18/1.44) affects the light yield value in a significant way. Number of photo-electrons  is lower for plastic bars by about 10$\%$-30$\%$.   

\newpage
\section{Activation analysis}\label{sec:activation}

The activation analysis performed in this section is similar to that performed for silicon photomultiplier arrays (SiPMs) in our previous publication \cite{POLAR-2_SiPMirradiation}. In analog way, two High Purity Germanium detectors (HPGe) with very good energy resolution were used to measure proton activation products for EJ-200 and EJ-248M. Only samples irradiated with the highest dose of 18.7~Gy were measured in this case. Both HPGe detectors were placed inside a low-background lead protected chambers to decrease background radiation. About fifteen minutes were required to move the irradiated samples from the experimental hall and place them into the HPGe setups (as for the SiPMs). This introduced a non-negligible time delay between the irradiation process and the start of data acquisitions, that limits our detection ability of decay products with decay times shorter than 3-4 minutes.

\begin{figure}[!h]
  \centering
  \includegraphics[scale=0.4]{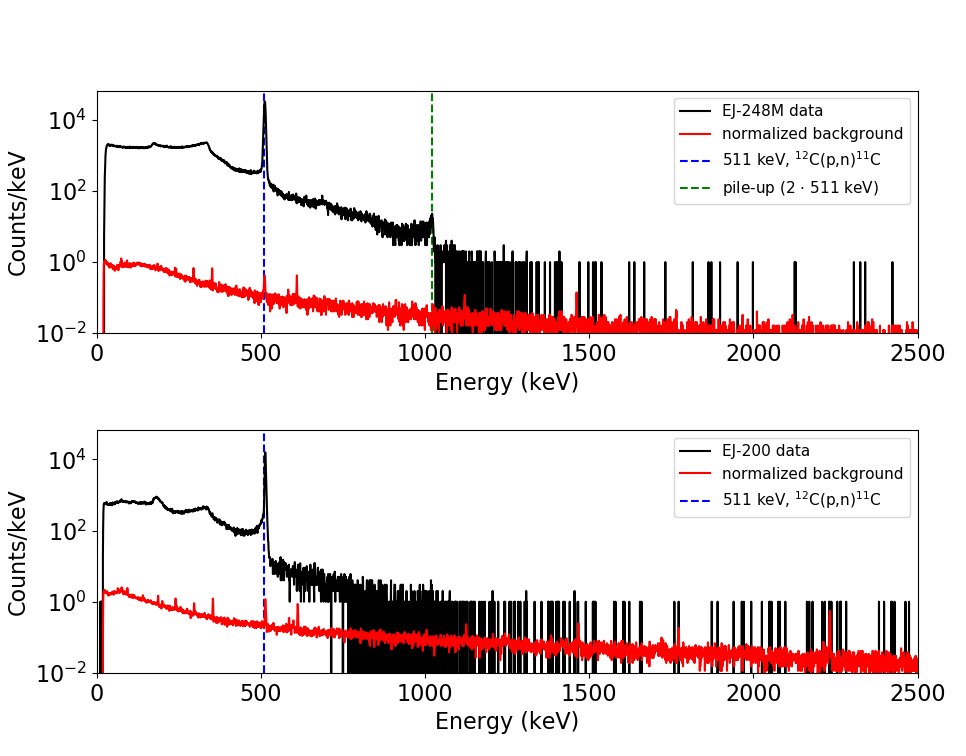}
  \caption{Gamma-ray energy spectrum measured with HPGe detectors after 58~MeV proton scintillators irradiation. Black lines show experimental data. Red lines present measured background normalized to the same live time for each HPGe detector. Dashed blue lines show the identified 511~keV peaks. The pile-up of this line was also observed at 1022~keV for one detector.}
  \label{fig:activ_energy}
\end{figure}

\begin{figure}[!h]
  \centering
  \includegraphics[scale=0.5]{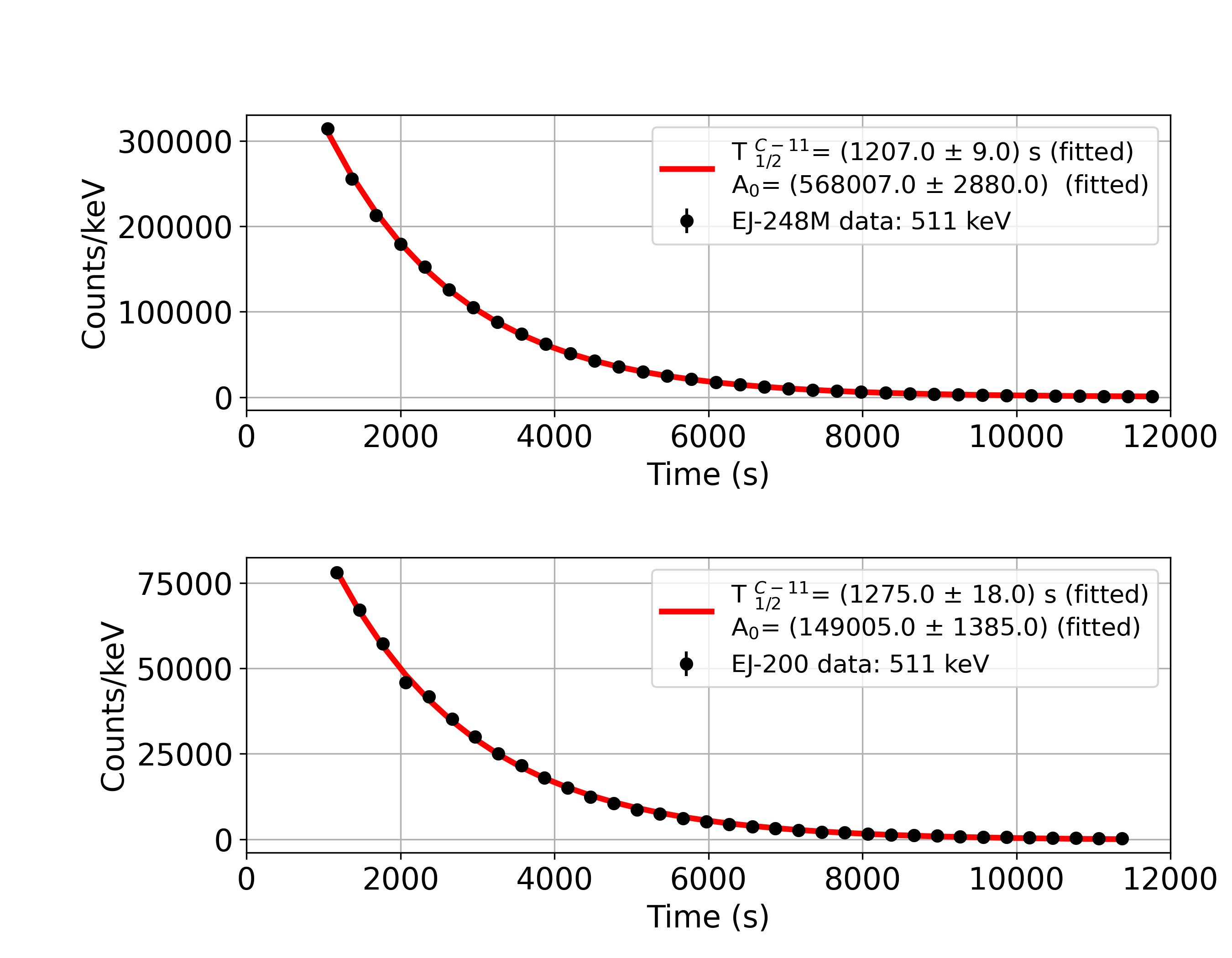}
  \caption{Decay time distributions of the 511~keV line (black points) measured for 58 MeV proton irradiated plastic scintillators with fitted exponential curves (red lines). The experimental uncertainties are smaller than the bullet marker size.}
  \label{fig:activ_time}
\end{figure}

Figure \ref{fig:activ_energy} shows protons induced gamma-ray spectra in the energy range up to 2.5~MeV for both scintillators. Only one 511~keV gamma-line from $^{12}$C(p,n)$^{11}$C reaction was observed, where daughter nuclei mostly decay into $\beta ^+$. This identification was confirmed by computing the decay time corresponding to this line. For this reason gamma-ray energy spectra were saved in a series (5~min intervals), where the number of counts for 511~keV line was determined for each file. Figure \ref{fig:activ_time} presents experimental data and fitted exponential curves of the 511~keV line decay. The obtained decay time values are respectively (1275$\pm$18)~s and (1207$\pm$9)~s for EJ-200 and EJ-248M. They are in good agreement (within 3$\sigma$) with the $\beta ^+$ decay time of 1221.83~s from $^{11}$C \cite{NuDat3}.

\section{Summary and outlook}

The POLAR-2 gamma-ray polarimeter exploits the low density of plastic scintillators in order to maximise the Compton scattering cross section down to a few keV. The Compton scattering angle distribution provide information about the polarization parameters. In this paper, we compare the performances of two types of plastic scintillators, namely EJ-200 and EJ-248M from Eljen Technology, under space-like radiation conditions.

Our results show similar spectral properties for the two scintillator types, including emission and absorption spectra as well as light yield. These properties were not significantly affected by the 58~MeV proton irradiation with several doses up to 18.7~Gy, that corresponds to about 80~years in low-Earth orbit (383~km altitude, 42$^\circ$ inclination) for a non-shielded scintillator. The starting value of emitted photo-electrons for all samples (500-600 p.e./476 keV) at a distance of 11~mm was reduced to about 30-40$\%$ at 118~mm, which is related to the light yield attenuation with increasing light distance inside the scintillator material. Obtained photo-electron numbers for a plastic bars are also lower comparing to the smaller cylindrical scintillator size.

Furthermore, a proton activation analysis was performed. The results show the presence of the 511~keV line for both scintillators in the energy spectrum measured with the HPGe detector. Observed line corresponds to $\beta ^+$ decay of $^{11}$C populated in proton reaction, for which the decay time is about 1222~s. This time being shorter than the CSS orbital period, we do not expect the activation products from the scintillators to provide a significant contribution in the polarimeter response degradation.

The radiation hardness of their optical properties makes both EJ-200 and EJ-248M well suited to be used for the POLAR-2 GRB polarimeter. Since the optical quality of the scintillators were tested to doses much higher than what we expect for POLAR-2, this conclusion also stands for longer life-time experiments, or for experiments with higher inclination/altitude whose orbit would be more exposed to space radiation.

Since both types of scintillators show similar degradation behavior under harsh radiation conditions, the choice between EJ-200 and EJ-248M for the final POLAR-2 design is mainly driven by their physical properties, like scintillator softening point and its surface polishing quality. The first factor suggests an advantage of EJ-248M, but the final decision will be made based on light transport simulations, that will take the second factor into consideration as well as the scintillator shape. Based on preliminary optical simulation results and optical light yield measurement in the lab, the EJ-248M type shows better performances for the POLAR-2 scintillator shape. This will be further studied in a future publication about POLAR-2's module optical characterization and simulation.

In this work, we have shown that the amount of light collected by the photodetector may be affected by the size and shape of the scintillators, and that plastic scintillators are highly resistant to space-like radiation conditions. The obtained data can be used to estimate the performance degradation of the two types of studied plastic scintillators after a given exposed dose. This can be useful to define the life-time of future experiments wanting to employ such scintillators in radiation environment. Performed activation analysis will also allow to determine the main contribution of induced radiation, which may affect the detector (as a whole system) and the data quality.

\newpage
\bmhead{Acknowledgments}
We gratefully acknowledge the Swiss Space Office of the State Secretariat for Education, Research and Innovation (ESA PRODEX Programme) which supported the development and production of the POLAR-2 detector. N.D.A. acknowledges the support of the Swiss National Science Foundation.
\newpage
\section*{Declarations}

\subsection*{Funding}

The (co-)authors are funded by the funding the agencies described in the acknowledgment section. 

\subsection*{Conflicts of Interest}
The authors declare that the research was conducted in the absence of any commercial or financial relationships that could be construed as a potential conflict of interest.

\subsection*{Consent to participate} 
Not applicable.

\subsection*{Consent for publication} 
Not applicable.

\subsection*{Code availability}
Not applicable.

\subsection*{Author's Contribution}
The main author is Slawomir Mianowski. The first draft of the manuscript was written by Slawomir Mianowski and Nicolas De Angelis, and all authors commented on previous versions of the manuscript.


\newpage
\bibliography{sn-bibliography}


\begin{thebibliography}{13}
\ifx \bisbn   \undefined \def \bisbn  #1{ISBN #1}\fi
\ifx \binits  \undefined \def \binits#1{#1}\fi
\ifx \bauthor  \undefined \def \bauthor#1{#1}\fi
\ifx \batitle  \undefined \def \batitle#1{#1}\fi
\ifx \bjtitle  \undefined \def \bjtitle#1{#1}\fi
\ifx \bvolume  \undefined \def \bvolume#1{\textbf{#1}}\fi
\ifx \byear  \undefined \def \byear#1{#1}\fi
\ifx \bissue  \undefined \def \bissue#1{#1}\fi
\ifx \bfpage  \undefined \def \bfpage#1{#1}\fi
\ifx \blpage  \undefined \def \blpage #1{#1}\fi
\ifx \burl  \undefined \def \burl#1{\textsf{#1}}\fi
\ifx \doiurl  \undefined \def \doiurl#1{\url{https://doi.org/#1}}\fi
\ifx \betal  \undefined \def \betal{\textit{et al.}}\fi
\ifx \binstitute  \undefined \def \binstitute#1{#1}\fi
\ifx \binstitutionaled  \undefined \def \binstitutionaled#1{#1}\fi
\ifx \bctitle  \undefined \def \bctitle#1{#1}\fi
\ifx \beditor  \undefined \def \beditor#1{#1}\fi
\ifx \bpublisher  \undefined \def \bpublisher#1{#1}\fi
\ifx \bbtitle  \undefined \def \bbtitle#1{#1}\fi
\ifx \bedition  \undefined \def \bedition#1{#1}\fi
\ifx \bseriesno  \undefined \def \bseriesno#1{#1}\fi
\ifx \blocation  \undefined \def \blocation#1{#1}\fi
\ifx \bsertitle  \undefined \def \bsertitle#1{#1}\fi
\ifx \bsnm \undefined \def \bsnm#1{#1}\fi
\ifx \bsuffix \undefined \def \bsuffix#1{#1}\fi
\ifx \bparticle \undefined \def \bparticle#1{#1}\fi
\ifx \barticle \undefined \def \barticle#1{#1}\fi
\bibcommenthead
\ifx \bconfdate \undefined \def \bconfdate #1{#1}\fi
\ifx \botherref \undefined \def \botherref #1{#1}\fi
\ifx \url \undefined \def \url#1{\textsf{#1}}\fi
\ifx \bchapter \undefined \def \bchapter#1{#1}\fi
\ifx \bbook \undefined \def \bbook#1{#1}\fi
\ifx \bcomment \undefined \def \bcomment#1{#1}\fi
\ifx \oauthor \undefined \def \oauthor#1{#1}\fi
\ifx \citeauthoryear \undefined \def \citeauthoryear#1{#1}\fi
\ifx \endbibitem  \undefined \def \endbibitem {}\fi
\ifx \bconflocation  \undefined \def \bconflocation#1{#1}\fi
\ifx \arxivurl  \undefined \def \arxivurl#1{\textsf{#1}}\fi
\csname PreBibitemsHook\endcsname

\bibitem[\protect\citeauthoryear{Mianowski
  et~al.}{2022}]{POLAR-2_SiPMirradiation}
\begin{barticle}
\bauthor{\bsnm{Mianowski}, \binits{S.}},
\bauthor{\bsnm{De~Angelis}, \binits{N.}},
\bauthor{\bsnm{Hulsman}, \binits{J.}},
\bauthor{\bsnm{Kole}, \binits{M.}},
\bauthor{\bsnm{Kowalski}, \binits{T.}},
\bauthor{\bsnm{Kusyk}, \binits{S.}},
\bauthor{\bsnm{Li}, \binits{H.}},
\bauthor{\bsnm{Mianowska}, \binits{Z.}},
\bauthor{\bsnm{Mietelski}, \binits{J.}},
\bauthor{\bsnm{Pollo}, \binits{A.}},
\bauthor{\bsnm{Rybka}, \binits{D.}},
\bauthor{\bsnm{Sun}, \binits{J.}},
\bauthor{\bsnm{Swakon}, \binits{J.}},
\bauthor{\bsnm{Wrobel}, \binits{D.}},
\bauthor{\bsnm{Wu}, \binits{X.}}:
\batitle{Proton irradiation of sipm arrays for polar-2}.
\bjtitle{Experimental Astronomy}
(\byear{2022})
\doiurl{10.1007/s10686-022-09873-6}
\end{barticle}
\endbibitem

\bibitem[\protect\citeauthoryear{Produit et~al.}{2018}]{ProduitPOLAR}
\begin{barticle}
\bauthor{\bsnm{Produit}, \binits{N.}},
\bauthor{\bsnm{Bao}, \binits{T.W.}},
\bauthor{\bsnm{Batsch}, \binits{T.}},
\bauthor{\bsnm{Bernasconi}, \binits{T.}},
\bauthor{\bsnm{Britvich}, \binits{I.}},
\bauthor{\bsnm{Cadoux}, \binits{F.}},
\bauthor{\bsnm{Cernuda}, \binits{I.}},
\bauthor{\bsnm{Chai}, \binits{J.Y.}},
\bauthor{\bsnm{Dong}, \binits{Y.W.}},
\bauthor{\bsnm{Gauvin}, \binits{N.}},
\bauthor{\bsnm{Hajdas}, \binits{W.}},
\bauthor{\bsnm{Kole}, \binits{M.}},
\bauthor{\bsnm{Kong}, \binits{M.N.}},
\bauthor{\bsnm{Kramert}, \binits{R.}},
\bauthor{\bsnm{Li}, \binits{L.}},
\bauthor{\bsnm{Liu}, \binits{J.T.}},
\bauthor{\bsnm{Liu}, \binits{X.}},
\bauthor{\bsnm{Marcinkowski}, \binits{R.}},
\bauthor{\bsnm{Orsi}, \binits{S.}},
\bauthor{\bsnm{Pohl}, \binits{M.}},
\bauthor{\bsnm{Rapin}, \binits{D.}},
\bauthor{\bsnm{Rybka}, \binits{D.}},
\bauthor{\bsnm{Rutczynska}, \binits{A.}},
\bauthor{\bsnm{Shi}, \binits{H.L.}},
\bauthor{\bsnm{Socha}, \binits{P.}},
\bauthor{\bsnm{Sun}, \binits{J.C.}},
\bauthor{\bsnm{Song}, \binits{L.M.}},
\bauthor{\bsnm{Szabelski}, \binits{J.}},
\bauthor{\bsnm{Traseira}, \binits{I.}},
\bauthor{\bsnm{Xiao}, \binits{H.L.}},
\bauthor{\bsnm{Wang}, \binits{R.J.}},
\bauthor{\bsnm{Wen}, \binits{X.}},
\bauthor{\bsnm{Wu}, \binits{B.B.}},
\bauthor{\bsnm{Zhang}, \binits{L.}},
\bauthor{\bsnm{Zhang}, \binits{L.Y.}},
\bauthor{\bsnm{Zhang}, \binits{S.N.}},
\bauthor{\bsnm{Zhang}, \binits{Y.J.}},
\bauthor{\bsnm{Zwolinska}, \binits{A.}}:
\batitle{Design and construction of the polar detector}.
\bjtitle{Nuclear Instruments and Methods in Physics Research Section A:
  Accelerators, Spectrometers, Detectors and Associated Equipment}
\bvolume{877},
\bfpage{259}--\blpage{268}
(\byear{2018})
\doiurl{10.1016/j.nima.2017.09.053}
\end{barticle}
\endbibitem

\bibitem[\protect\citeauthoryear{De~Angelis et~al.}{2021}]{ICRC21_POLAR-2_NDA}
\begin{barticle}
\bauthor{\bsnm{De~Angelis}, \binits{N.}}, \betal:
\batitle{{Development and science perspectives of the POLAR-2 instrument: a
  large scale GRB polarimeter}}.
\bjtitle{PoS}
\bvolume{ICRC2021},
\bfpage{580}
(\byear{2021})
\doiurl{10.22323/1.395.0580}
\end{barticle}
\endbibitem

\bibitem[\protect\citeauthoryear{}{}]{Eljen}
\begin{botherref}
Eljen Technology.
\url{https://eljentechnology.com/products/plastic-scintillators}
\end{botherref}
\endbibitem

\bibitem[\protect\citeauthoryear{Kole et~al.}{2021}]{ICRC21_POLAR-2_MK}
\begin{barticle}
\bauthor{\bsnm{Kole}, \binits{M.}}, \betal:
\batitle{{Gamma-Ray Polarization Results of the POLAR Mission and Future
  Prospects}}.
\bjtitle{PoS}
\bvolume{ICRC2021},
\bfpage{600}
(\byear{2021})
\doiurl{10.22323/1.395.0600}
\end{barticle}
\endbibitem

\bibitem[\protect\citeauthoryear{Michalec et~al.}{2010}]{MICHALEC2010738}
\begin{barticle}
\bauthor{\bsnm{Michalec}, \binits{B.}},
\bauthor{\bsnm{Swakon}, \binits{J.}},
\bauthor{\bsnm{Sowa}, \binits{U.}},
\bauthor{\bsnm{Ptaszkiewicz}, \binits{M.}},
\bauthor{\bsnm{Cywicka-Jakiel}, \binits{T.}},
\bauthor{\bsnm{Olko}, \binits{P.}}:
\batitle{Proton radiotherapy facility for ocular tumors at the ifj pan in
  krakow poland}.
\bjtitle{Applied Radiation and Isotopes}
\bvolume{68},
\bfpage{738}--\blpage{742}
(\byear{2010})
\doiurl{10.1016/j.apradiso.2009.11.001}
\end{barticle}
\endbibitem

\bibitem[\protect\citeauthoryear{{De Angelis} et~al.}{2023}]{POLAR-2Annealing}
\begin{barticle}
\bauthor{\bsnm{{De Angelis}}, \binits{N.}},
\bauthor{\bsnm{Kole}, \binits{M.}},
\bauthor{\bsnm{Cadoux}, \binits{F.}},
\bauthor{\bsnm{Hulsman}, \binits{J.}},
\bauthor{\bsnm{Kowalski}, \binits{T.}},
\bauthor{\bsnm{Kusyk}, \binits{S.}},
\bauthor{\bsnm{Mianowski}, \binits{S.}},
\bauthor{\bsnm{Rybka}, \binits{D.}},
\bauthor{\bsnm{Stauffer}, \binits{J.}},
\bauthor{\bsnm{Swakon}, \binits{J.}},
\bauthor{\bsnm{Wrobel}, \binits{D.}},
\bauthor{\bsnm{Wu}, \binits{X.}}:
\batitle{Temperature dependence of radiation damage annealing of silicon
  photomultipliers}.
\bjtitle{Nuclear Instruments and Methods in Physics Research Section A:
  Accelerators, Spectrometers, Detectors and Associated Equipment}
\bvolume{1048},
\bfpage{167934}
(\byear{2023})
\doiurl{10.1016/j.nima.2022.167934}
\end{barticle}
\endbibitem

\bibitem[\protect\citeauthoryear{Cumani et~al.}{2019}]{leobackground_2019}
\begin{barticle}
\bauthor{\bsnm{Cumani}, \binits{P.}},
\bauthor{\bsnm{Hernanz}, \binits{M.}},
\bauthor{\bsnm{Kiener}, \binits{J.}},
\bauthor{\bsnm{Tatischeff}, \binits{V.}},
\bauthor{\bsnm{Zoglauer}, \binits{A.}}:
\batitle{Background for a gamma-ray satellite on a low-earth orbit}.
\bjtitle{Experimental Astronomy}
\bvolume{47}(\bissue{3}),
\bfpage{273}--\blpage{302}
(\byear{2019})
\doiurl{10.1007/s10686-019-09624-0}
\end{barticle}
\endbibitem

\bibitem[\protect\citeauthoryear{Daly}{1989}]{1989AIPC..186..483D}
\begin{bchapter}
\bauthor{\bsnm{Daly}, \binits{E.}}:
\bctitle{Radiation environment evaluation for esa projects}.
In: \beditor{\bsnm{C.}, \binits{J.R.A.}},
\beditor{\bsnm{{Trombka}}, \binits{J.I.}} (eds.)
\bbtitle{High-Energy Radiation Background in Space},
vol. \bseriesno{186},
pp. \bfpage{483}--\blpage{499}
(\byear{1989})
\end{bchapter}
\endbibitem

\bibitem[\protect\citeauthoryear{Agostinelli et~al.}{2003}]{AGOSTINELLI2003250}
\begin{barticle}
\bauthor{\bsnm{Agostinelli}, \binits{S.}},
\bauthor{\bsnm{Allison}, \binits{J.}},
\bauthor{\bsnm{Amako}, \binits{K.}},
\bauthor{\bsnm{Apostolakis}, \binits{J.}},
\bauthor{\bsnm{Araujo}, \binits{H.}},
\bauthor{\bsnm{Arce}, \binits{P.}},
\bauthor{\bsnm{Asai}, \binits{M.}},
\bauthor{\bsnm{Axen}, \binits{D.}},
\bauthor{\bsnm{Banerjee}, \binits{S.}},
\bauthor{\bsnm{Barrand}, \binits{G.}},
\bauthor{\bsnm{Behner}, \binits{F.}},
\bauthor{\bsnm{Bellagamba}, \binits{L.}},
\bauthor{\bsnm{Boudreau}, \binits{J.}},
\bauthor{\bsnm{Broglia}, \binits{L.}},
\bauthor{\bsnm{Brunengo}, \binits{A.}},
\bauthor{\bsnm{Burkhardt}, \binits{H.}},
\bauthor{\bsnm{Chauvie}, \binits{S.}},
\bauthor{\bsnm{Chuma}, \binits{J.}},
\bauthor{\bsnm{Chytracek}, \binits{R.}},
\bauthor{\bsnm{Cooperman}, \binits{G.}},
\bauthor{\bsnm{Cosmo}, \binits{G.}},
\bauthor{\bsnm{Degtyarenko}, \binits{P.}},
\bauthor{\bsnm{Dell'Acqua}, \binits{A.}},
\bauthor{\bsnm{Depaola}, \binits{G.}},
\bauthor{\bsnm{Dietrich}, \binits{D.}},
\bauthor{\bsnm{Enami}, \binits{R.}},
\bauthor{\bsnm{Feliciello}, \binits{A.}},
\bauthor{\bsnm{Ferguson}, \binits{C.}},
\bauthor{\bsnm{Fesefeldt}, \binits{H.}},
\bauthor{\bsnm{Folger}, \binits{G.}},
\bauthor{\bsnm{Foppiano}, \binits{F.}},
\bauthor{\bsnm{Forti}, \binits{A.}},
\bauthor{\bsnm{Garelli}, \binits{S.}},
\bauthor{\bsnm{Giani}, \binits{S.}},
\bauthor{\bsnm{Giannitrapani}, \binits{R.}},
\bauthor{\bsnm{Gibin}, \binits{D.}},
\bauthor{\bsnm{{Gómez Cadenas}}, \binits{J.J.}},
\bauthor{\bsnm{González}, \binits{I.}},
\bauthor{\bsnm{{Gracia Abril}}, \binits{G.}},
\bauthor{\bsnm{Greeniaus}, \binits{G.}},
\bauthor{\bsnm{Greiner}, \binits{W.}},
\bauthor{\bsnm{Grichine}, \binits{V.}},
\bauthor{\bsnm{Grossheim}, \binits{A.}},
\bauthor{\bsnm{Guatelli}, \binits{S.}},
\bauthor{\bsnm{Gumplinger}, \binits{P.}},
\bauthor{\bsnm{Hamatsu}, \binits{R.}},
\bauthor{\bsnm{Hashimoto}, \binits{K.}},
\bauthor{\bsnm{Hasui}, \binits{H.}},
\bauthor{\bsnm{Heikkinen}, \binits{A.}},
\bauthor{\bsnm{Howard}, \binits{A.}},
\bauthor{\bsnm{Ivanchenko}, \binits{V.}},
\bauthor{\bsnm{Johnson}, \binits{A.}},
\bauthor{\bsnm{Jones}, \binits{F.W.}},
\bauthor{\bsnm{Kallenbach}, \binits{J.}},
\bauthor{\bsnm{Kanaya}, \binits{N.}},
\bauthor{\bsnm{Kawabata}, \binits{M.}},
\bauthor{\bsnm{Kawabata}, \binits{Y.}},
\bauthor{\bsnm{Kawaguti}, \binits{M.}},
\bauthor{\bsnm{Kelner}, \binits{S.}},
\bauthor{\bsnm{Kent}, \binits{P.}},
\bauthor{\bsnm{Kimura}, \binits{A.}},
\bauthor{\bsnm{Kodama}, \binits{T.}},
\bauthor{\bsnm{Kokoulin}, \binits{R.}},
\bauthor{\bsnm{Kossov}, \binits{M.}},
\bauthor{\bsnm{Kurashige}, \binits{H.}},
\bauthor{\bsnm{Lamanna}, \binits{E.}},
\bauthor{\bsnm{Lampén}, \binits{T.}},
\bauthor{\bsnm{Lara}, \binits{V.}},
\bauthor{\bsnm{Lefebure}, \binits{V.}},
\bauthor{\bsnm{Lei}, \binits{F.}},
\bauthor{\bsnm{Liendl}, \binits{M.}},
\bauthor{\bsnm{Lockman}, \binits{W.}},
\bauthor{\bsnm{Longo}, \binits{F.}},
\bauthor{\bsnm{Magni}, \binits{S.}},
\bauthor{\bsnm{Maire}, \binits{M.}},
\bauthor{\bsnm{Medernach}, \binits{E.}},
\bauthor{\bsnm{Minamimoto}, \binits{K.}},
\bauthor{\bsnm{{Mora de Freitas}}, \binits{P.}},
\bauthor{\bsnm{Morita}, \binits{Y.}},
\bauthor{\bsnm{Murakami}, \binits{K.}},
\bauthor{\bsnm{Nagamatu}, \binits{M.}},
\bauthor{\bsnm{Nartallo}, \binits{R.}},
\bauthor{\bsnm{Nieminen}, \binits{P.}},
\bauthor{\bsnm{Nishimura}, \binits{T.}},
\bauthor{\bsnm{Ohtsubo}, \binits{K.}},
\bauthor{\bsnm{Okamura}, \binits{M.}},
\bauthor{\bsnm{O'Neale}, \binits{S.}},
\bauthor{\bsnm{Oohata}, \binits{Y.}},
\bauthor{\bsnm{Paech}, \binits{K.}},
\bauthor{\bsnm{Perl}, \binits{J.}},
\bauthor{\bsnm{Pfeiffer}, \binits{A.}},
\bauthor{\bsnm{Pia}, \binits{M.G.}},
\bauthor{\bsnm{Ranjard}, \binits{F.}},
\bauthor{\bsnm{Rybin}, \binits{A.}},
\bauthor{\bsnm{Sadilov}, \binits{S.}},
\bauthor{\bsnm{{Di Salvo}}, \binits{E.}},
\bauthor{\bsnm{Santin}, \binits{G.}},
\bauthor{\bsnm{Sasaki}, \binits{T.}},
\bauthor{\bsnm{Savvas}, \binits{N.}},
\bauthor{\bsnm{Sawada}, \binits{Y.}},
\bauthor{\bsnm{Scherer}, \binits{S.}},
\bauthor{\bsnm{Sei}, \binits{S.}},
\bauthor{\bsnm{Sirotenko}, \binits{V.}},
\bauthor{\bsnm{Smith}, \binits{D.}},
\bauthor{\bsnm{Starkov}, \binits{N.}},
\bauthor{\bsnm{Stoecker}, \binits{H.}},
\bauthor{\bsnm{Sulkimo}, \binits{J.}},
\bauthor{\bsnm{Takahata}, \binits{M.}},
\bauthor{\bsnm{Tanaka}, \binits{S.}},
\bauthor{\bsnm{Tcherniaev}, \binits{E.}},
\bauthor{\bsnm{{Safai Tehrani}}, \binits{E.}},
\bauthor{\bsnm{Tropeano}, \binits{M.}},
\bauthor{\bsnm{Truscott}, \binits{P.}},
\bauthor{\bsnm{Uno}, \binits{H.}},
\bauthor{\bsnm{Urban}, \binits{L.}},
\bauthor{\bsnm{Urban}, \binits{P.}},
\bauthor{\bsnm{Verderi}, \binits{M.}},
\bauthor{\bsnm{Walkden}, \binits{A.}},
\bauthor{\bsnm{Wander}, \binits{W.}},
\bauthor{\bsnm{Weber}, \binits{H.}},
\bauthor{\bsnm{Wellisch}, \binits{J.P.}},
\bauthor{\bsnm{Wenaus}, \binits{T.}},
\bauthor{\bsnm{Williams}, \binits{D.C.}},
\bauthor{\bsnm{Wright}, \binits{D.}},
\bauthor{\bsnm{Yamada}, \binits{T.}},
\bauthor{\bsnm{Yoshida}, \binits{H.}},
\bauthor{\bsnm{Zschiesche}, \binits{D.}}:
\batitle{Geant4—a simulation toolkit}.
\bjtitle{Nuclear Instruments and Methods in Physics Research Section A:
  Accelerators, Spectrometers, Detectors and Associated Equipment}
\bvolume{506},
\bfpage{250}--\blpage{303}
(\byear{2003})
\doiurl{10.1016/S0168-9002(03)01368-8}
\end{barticle}
\endbibitem

\bibitem[\protect\citeauthoryear{Swiderski et~al.}{2010}]{Swiderski}
\begin{barticle}
\bauthor{\bsnm{Swiderski}, \binits{L.}},
\bauthor{\bsnm{Moszynski}, \binits{M.}},
\bauthor{\bsnm{Czarnacki}, \binits{W.}},
\bauthor{\bsnm{Iwanowska}, \binits{J.}},
\bauthor{\bsnm{Syntfeld-Kazuch}, \binits{A.}},
\bauthor{\bsnm{Szczesniak}, \binits{T.}},
\bauthor{\bsnm{Pausch}, \binits{G.}},
\bauthor{\bsnm{Plettner}, \binits{C.}},
\bauthor{\bsnm{Roemer}, \binits{K.}}:
\batitle{Measurement of compton edge position in low-z scintillators}.
\bjtitle{Radiation Measurements}
\bvolume{45},
\bfpage{605}--\blpage{607}
(\byear{2010})
\doiurl{10.1016/j.apradiso.2009.11.001}
\end{barticle}
\endbibitem

\bibitem[\protect\citeauthoryear{}{}]{POLAR2Optsim}
\begin{botherref}
De Angelis, N.: Optical simulations and characterization of a POLAR-2
  polarimeter module. \textit{In preparation} (2023)
\end{botherref}
\endbibitem

\bibitem[\protect\citeauthoryear{}{}]{NuDat3}
\begin{botherref}
The National Nuclear Data Center at Brookhaven National Laboratory.
\url{https://www.nndc.bnl.gov/nudat3/}
\end{botherref}
\endbibitem

\end{thebibliography}


\end{document}